# Photon number squeezing of ultra-broadband laser pulses generated by microstructure fibers

**Kenichi Hirosawa, Hiroto Furumochi, Atsushi Tada, and Fumihiko Kannari**
*Department of Electronics and Electrical Engineering, Keio University, 3-14-1 Hiyoshi, Kohoku-ku, Yokohama, 223-8522, Japan*

**Masahiro Takeoka and Masahide Sasaki**
*National Institute of Information and Communications Technology, 4-2-1 Nukui-kitamachi, Koganei, Tokyo, 184-8795, Japan*

To the best of our knowledge, we demonstrate for the first time the generation of photon number squeezing by spectral filtering for ultra-broadband light generated by microstructure fibers at 800 nm. A maximum squeezing of 4.6 dB is observed, corresponding to 10.3 dB after correcting for detection losses. We numerically analyzed the quantum dynamics of ultrashort laser pulse propagation through optical fibers by solving a nonlinear quantum Schrödinger equation that included Raman scattering, especially for the quantum correlation of photon number fluctuation among frequency modes in broadband pulses.

The generation of squeezed light in optical fibers has been attracting considerable attention for creating and distributing continuous variable entanglements for quantum communication applications. It was recently demonstrated that more than 50% photon number squeezing is possible with conventional glass fibers and spectral filtering at a wavelength of 1.5μm[1]. In such soliton propagation, squeezing is mainly induced by group-velocity dispersion (GVD) and self-phase modulation (SPM). In fact, for pulses with sufficient peak intensity the nonlinear effects are enough for observation of sub-Poissonian noise, even at the regime of normal GVD without the assistance of soliton propagation[2]. With high input peak intensities or long fiber lengths at the regime of anomalous dispersion, Raman self-frequency shift appears. Although it was reported that greater squeezing was obtained with Raman self-frequency shifted soliton[3], it generally does not enhance squeezing because it is physically connected with noise from stimulated Raman scattering (SRS). Numerical model simulations confirmed asymmetric squeezing behavior for low and high-pass spectral filtering in the presence of SRS, which agrees well with the experiments[3], resulting in less squeezing compared to numerical calculations that ignore the Raman effect[4]. Therefore, the net effect of SRS is a deterioration of squeezing.

Recent experiments with microstructure fibers (MFs) have demonstrated that one can achieve dramatic nonlinear effects even in short lengths of fiber. Because of their unusual GVD characteristics, one can realize zero-GVD pulse propagation in visible and near-infrared regions. It is well-known that extremely broad spectra can be generated when ultrashort pulses at near-zero GVD wavelength are launched into MFs. In addition to self-phase modulation (SPM)-induced spectral broadening, SRS and parametric generation play a crucial role in ultra-broadband light generation. Four-wave mixing (FWM) processes can be phase matched near-zero GVD, significantly enhancing spectral broadening[5]. So far, two experiments have already reported the generation of squeezed light with MFs at 1.5 μm[6] and 800 nm[7]. Properly designed MFs could support soliton pulse propagation at lower pulse energies, even at 800 nm. However, the unique features of MFs have not increased squeezed light generation benefits. In this Letter we report, for the first time to our knowledge, the generation of squeezed light by spectrally filtering ultra-broadband pulses generated from MFs. Photon number squeezing was observed around Raman self-shifting components. We also numerically analyzed the formation of quantum correlation among frequency modes in the ultra-broad spectrum by solving a nonlinear quantum Schrödinger equation.

The MF used in our experiments exhibited zero GVD at 820 nm consisting of an ~3.06-μm diameter core surrounded by a hexagonal array of ~1.74 μm diameter air voids. Figure 1 shows the output spectra from a 30 cm MF for various launched pulse energies. An input laser pulse is obtained from a mode-locked Ti: sapphire laser oscillator whose spectrum width is ~45 nm (FWHM). The residual dispersion originated from the oscillator, and optics was compensated for with a programmable pulse shaper located between the oscillator and the MF. Thus, the launched pulse is a transform limited pulse whose center wavelength was chosen at 810 nm. The output spectra exhibit Raman self-shifting, the generation of anti-Stokes Raman, and other FWMs. The mechanism of continuum generation is well understood by solving a classical nonlinear Schrödinger equation. The anti-Stokes Raman component is trapped by the Stokes pulse and co-propagates with spectral shift toward shorter wavelengths as the Raman spectrum shifts toward longer wavelengths[8]. This mechanism involving cross-phase modulation between the anti-Stokes and Stokes pulses may cause some quantum correlation between them. Higher launched pulse energies generate wider output spectra. The output light was collimated through a lens and angularly dispersed by a blazed, gold-coated grating. The total reflector forms a spectral Fourier transform plane on which we placed a knife-edge filter to act as a spectral filter. The filtered beam was then focused through a 50:50 beam splitter consisting of a half-wave plate and a polarization beam splitter onto two silicon photodiodes, which served as a balanced detection system. The sum and difference AC photocurrents of the two detectors were recorded by an RF spectrum analyzer at 10 MHz with a bandwidth of 100 kHz, a video bandwidth of 100 Hz that averaged over 100 measurement cycles after an RF amplifier, and a low-pass filter. In balanced detection, due to the limited bandwidth of optical elements, we detected only a partial spectrum, ranging from 750-950 nm. The estimated measured detection efficiency is 75% at 800 nm. The two detectors were balanced to more than -20dB extinction ratio.

Figure 2 (a) shows plots of quantum noise reduction for various cutoff wavelengths of the low-pass filter. The launched laser energy is 120 pJ. When the spectral band of 950-850 nm was preserved, the maximum photon number squeezing of 4.6dB was observed, which corresponds to 10.3dB after correcting for detection losses.  Note that the maximum photon number squeezing obtained during our separate experiment, which used the same laser pulse and experimental setup with a conventional polarization maintained glass fiber, is only ~0.4dB at 800 nm. The longest spectral peak around 915 nm corresponds to a Raman self-frequency shift, which gradually shifts toward longer wavelengths during propagation. This stimulated Raman pulse exhibits almost soliton-like behavior because its wavelength has already reached the anomalous GVD region of this MF. It is surprising that squeezing was observed when the entire Raman component is preserved with a low-pass filter. It is different from previous soliton pulse squeezing experiments at 1.5 μm with conventional glass fibers, where negative quantum correlation was observed for only a fraction of Raman self-shifting spectrum with a high-pass filter. Therefore, our observation is inconsistent with our speculation that SRS noise prevents the soliton pulse from establishing a negative quantum correlation in the photon number inside the Raman spectrum. It is the first observation that photon number squeezing is obtainable around a Raman component that separately propagates from the main pulses.

Figure 2(b) shows similar plots of quantum noise reduction for various high-pass filters. The launched laser energy was 112 pJ. When the spectral band of 850-750 nm was preserved, photon number squeezing of 3.5dB was observed. Quantum noise reduction monotonously decreased as launched laser energy decreased. At 63 pJ, we did not obtain squeezing with any low-pass filters, although squeezing of ~0.7dB was still obtainable with high-pass filters. The filter edge wavelength, which caused the highest noise reduction, shifts toward longer wavelength as the Raman component shifts.

Despite various optical nonlinear processes involved in the generation of such broadband pulses, quantum correlation can be established among frequency modes. We carried out a numerical simulation to ensure the occurrence of photon number squeezing in ultra-broadband pulses generated from MFs and investigated the physics. Propagation of a normalized field amplitude operator in a fiber with Kerr nonlinearity is described by a quantum nonlinear Schrödinger equation. To analyze the evolution of quantum correlation in the presence of Raman self-shifting, SRS was included in the equation using a Raman response function[9]. Up to a 5th order dispersion was considered. For any given initial pulse, numerical calculation of the Fano factor was performed by a back propagation method[10]. We calculated the intrapulse cross-correlation[11] of the photon number fluctuation to observe the intermode quantum structure.

When SRS is added to numerical analysis for conventional soliton propagation at 1.5 μm, the negative quantum correlation between the longer spectral wing and the spectral center is enhanced, whereas the negative quantum correlation tends to be lost between the shorter wing and the spectral center. Moreover, Raman self-shifting components exhibit a stronger positive quantum correlation. Figure 3 shows the numerically obtained output spectrum from a 30 cm MF and the intrapulse quantum correlations. The launched energy is 118 pJ with a pulse width of 38 fs (FWHM). Since the dispersion curve of the MF used in our experiment was available only near the zero-GVD wavelength, we extrapolated the dispersion curve over the entire broad frequency region that we observed. However, due to the artificially formed dispersion curve, the predicted spectrum is narrower than that obtained in our experiment. Particularly, the spectral shift toward the shorter wavelength is small. Also, the center wavelength around 810 nm is more suppressed than the experimental results. Although more conditioning on the fiber parameters is necessary to reproduce the experimental spectra, it is still useful to observe intra-pulse quantum correlations for this spectrum that clearly exhibit

Raman self-shifting and corresponding anti-Stokes pulse generation. Although the quantum correlation is relatively strong and positive inside the Raman spectrum, the negative correlation is established between the Raman wavelength and some shorter spectral components. A strong negative correlation also exists around the anti-Stokes spectrum. Figure 3 shows plots of relative noise reduction obtained for low- and high-pass filters. Although the predicted noise reduction is around 2dB, which is smaller than our experiments, photon number squeezing is predicted for both low- and high-pass filters. As observed in our experiment, squeezing can be obtained when the entire Raman spectrum is preserved by low-pass filters. The spectral window where the squeezing is obtained is much wider than our experimental result in Fig. 2. Actually, we obtained relatively wide windows at lower launched energies. Therefore, presumably the nonlinearity of the MF estimated in our model is lower than the experiment.

Recently, we proposed an adaptive waveform control approach to construct designed quantum correlations in the spectrum of a femtosecond laser pulse by a nonlinear fiber spectral filtering method[12]. The controllability of the quantum correlation in the ultra-broadband pulse generated from MFs by manipulating the launched laser pulse shape and the properties of MFs in nonlinearity and dispersion need more study.   Ultra-broadband spectra may expand the frequency resources that can be quantum mechanically controlled by the classical method.

In conclusion, we have experimentally demonstrated that photon number squeezing can be observed around Raman self-shifting components in ultra-broadband pulses generated from MFs. By preserving the entire Raman component with a low-pass filter, we obtained a maximum photon number squeezing of 4.6 dB at 800 nm. The numerical mode qualitatively verified our experimental observations.

The present research was supported by Grants-in-Aid for Scientific Research (14350191) and by Grants-in-Aid for the 21th Century COE program for Optical and Electrical Technology from the Ministry of Education, Culture, Sports, Science and Technology, Japan.

**Reference**

1. S. R. Friberg, S. Machida, M. J. Werner, A. Levanon, and T. Mukai, Phys. Rev. Lett. **77**, 3775 (1996).
2. F. König, S. Spälter, I. L. Shumay, A. Sizmann, Th. Fauster, and G. Leuchs, J. Mod. Opt. **45**, 2425 (1998).
3. S. Spälter, M. Burk, U. Strößner, A. Sizmann, and G. Leuchs, Opt. Express **2**, 77 (1998).
4. M. J. Werner, Phys. Rev. **A 60**, 781 (1999).
5. J. M. Dudley, and S. Coen, IEEE J. Sel. Top. Quantum Electron. **8**, 651 (2002).
6. M. Fiorentino, J. E. Sharping, P. Kumar, D. Levandovsky, and M. Vasilyev, Opt. Lett. **27**, 649 (2002).
7. S. Lorentz, Ch. Silberhorn, N. Korolkova, R. S. Windeler, and G. Leuchs, Appl. Phys. **B 73**, 855 (2001).
8. N. Nishizawa, and T. Goto, Opt. Lett. **27**, 152 (2002).
9. D. Hollenbeck, and C. D. Cantrell, J. Opt. Soc. Am. **B 19**, 2886 (2002).
10. Y. Lai, and S. –S. Yu, Phys. Rev. **A 51**, 817 (1995).
11. S. Spälter, N. Korolkova, F. König, A. Sizmann, and G. Leuchs, Phys. Rev. Lett. **81**, 786 (1998).
12. M. Takeoka, D. Fujishima, and F. Kannari, Opt. Lett. **26**, 1592 (2001).

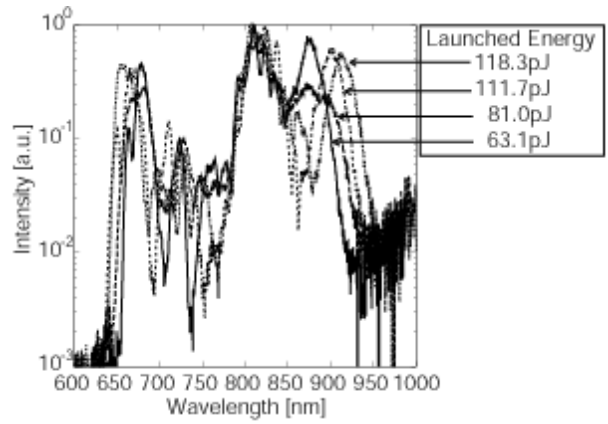

Fig. 1. Output spectra from a 30cm MF for various launched pulse energies.

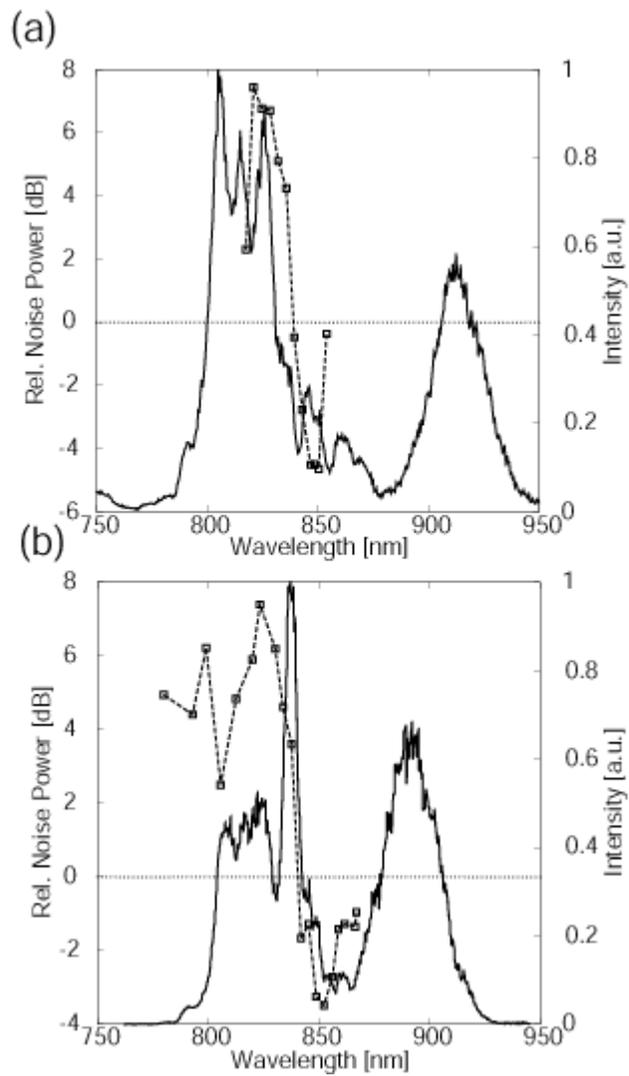

Fig. 2. Quantum noise reduction (relative to shot noise) for various cut-off wavelengths of (a) low-pass filtering at a launched energy of 118.3pJ, and (b) high-pass filtering at 111.7pJ. Output pulse spectra before filtering are shown.

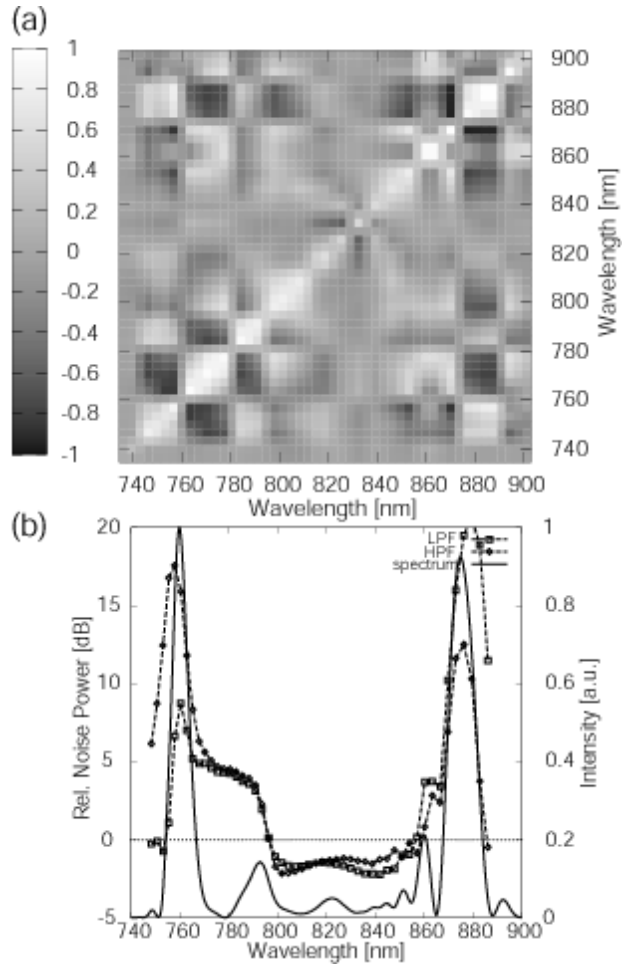

Fig. 3. (a) A numerically obtained map of intra-pulse quantum correlations. (b) An output spectrum (Solid line) before filtering and quantum noise reduction for various low- (circle plots) and high-pass (square plots) filtering.